\documentclass[12pt,a4paper,oneside]{article}
\usepackage{hyperref}
\usepackage{geometry}

\usepackage{indentfirst}
\renewcommand\thesection{\Roman{section}} 
\usepackage{abstract}
\usepackage{caption}
\usepackage{array}
\usepackage{booktabs}
\usepackage{titling}
\usepackage{titlesec}
\usepackage{graphicx}
\usepackage[mathscr]{eucal}
\usepackage{amsmath}
\usepackage{booktabs}
\usepackage{diagbox}
\usepackage{slashed}
\usepackage{authblk}
\usepackage{xcolor}
\usepackage{cite}
\titleformat{\section}[block]{\large\scshape}{\thesection.}{1em}{}
\title{Decay properties of $D_{s0}^*(2317)^+$ as a conventional $c\bar s$ meson}

\author{Meng Han\thanks{20208015011@stumail.hbu.edu.cn}, Wei Li, Su-Yan Pei, Ting-Ting Liu, Guo-Li Wang\thanks {wgl@hbu.edu.cn, corresponding author}}
\affil{Department of Physics and Technology,
Hebei University, Baoding, 071002, China}
\affil{Hebei Key Laboratory of High-Precision Computation and Application
of Quantum Field Theory, Baoding,071002, China}
\affil{Research Center for Computational Physics of Hebei Province, Baoding,071002, China}
\date{}

\begin{document}
\captionsetup[figure]{labelfont={bf},labelformat={default},labelsep=period,name={Fig.}}
\maketitle

\centerline{\large Abstract}

Taking $D_{s0}^{*}(2317)^{+ }$ as a conventional $c\bar s$ meson, we calculate its dominant strong and electromagnetic decays in the framework of the Bethe-Salpeter method. Our results are $\Gamma(D_{s0}^{*+}\to D_s^+\pi^0) = 7.83^{+1.97}_{-1.55}$ keV and $\Gamma(D_{s0}^{*+}\to D_s^{*+}\gamma) = 2.55^{+0.37}_{-0.45}$ keV.
The contributions of the different partial waves from the initial and final state wave functions to the decay width are also calculated, and we find that the relativistic corrections in both decay processes are very large.
\newpage
\section{Introduction}

The particle $D_{s0}^*(2317)^+$ was discovered in the invariant mass distribution of $D_s^+\pi^0$ by the BABAR collaboration in 2003 \cite{PRL90}. According to an analysis of experimental data, it is proposed that its isospin and spin parity quantum number are $I(J^P)=0 (0^+)$, its mass is $M$=2317.8 $\pm$ 0.5 MeV, and its full width is $\Gamma < 3.8$ MeV \cite{PTEP}. Since its discovery, its small width and low mass \cite{PLA19} have aroused great interest among experimentalists \cite{PRL91,PRD6803,PRL92} and high-energy physics theorists \cite{PLB568,PRD68,PLB570,PRD69,PRD72,PRD73,PLB566,PRL93,PRD7207,PLB567,PRD6805,PLB582,PA733,PRL9101,PLB578,PRD7003}.  Experts have not fully understood the internal structure of $D_{s0}^*(2317)$; it has been interpreted as a traditional $c\bar{s}$ state \cite{PLB568,PRD68,PLB570,PRD69,PRD72,PRD73}, while others think it might be an exotic meson state, such as a D-K bound state  \cite{PRD6805,PLB582,PA733,gavin21}, ${c}\bar {s}q\bar{q}$ teraquark state \cite{PLB566,PRL93,PRD7207}, or $D_s\pi$ quasibound state \cite{PLB567} or a mixture of a $c\bar{s}$ meson and a ${c}\bar{s}q\bar{q}$ tetraquark \cite{PLB578,PRD7003}, $c\bar{s}$ meson combined with quark-antiquark and meson-meson interpolators \cite{Mohler:2013rwa}, $c\bar{s}$ meson whith the effect of nearby $DK$ threshold taken into account by employing the corresponding four-quark operators \cite{Bali:2017pdv}, $c\bar{s}$ meson using interpolating fields of different structures \cite{alexandrou} etc. A more detailed description of exotic hadrons can be found in Ref.\cite{guo2023}.

In order to determine its internal structure, futher study of its production and decay is needed, especially its decay, which is important for revealing its internal composition, because different internal structures will lead to different decay behaviors and branching ratios. If it is a traditional meson, the  $D_{s0}^*(2317)$ mass is lower than the threshold of $D_s+\eta$ or $D_{u,d}+K$; thus, these possible Okubo-Zweig-Iizuka(OZI)-allowed decay channels  are kinematically forbidden.
Cho and Wise proposed the $\eta-\pi^0$ mixing mechanism and the origin of isospin violating effects from the mass splitting of $u$ and $d$ quarks \cite{PRD49}. In this case, $D_{s0}^*(2317)^+$ is converted first into $D_s^+\eta$ and then into $D_s^+\pi^0$ by the mixing, $i.e.$, $D_{s0}^{*}(2317)^{+}\to D_{s}^{+}\eta \to D_{s}^{+}\pi ^{0}$. Therefore, the strong decay channel $D_{s0}^{*}(2317)^{+}\to D_{s}^{+}\pi ^{0}$ is the dominant decay channel of this particle. This conclusion is supported by experimental data \cite{PTEP}.
$D_{s0}^*(2317)$ has another permissible decay channel, $i.e.$, the electromagnetic (EM) decay channel $D_{s0}^*(2317)\rightarrow D_{s}^{*}\gamma $.   Normally, EM decay is far narrower than strong decay, but because it is not affected by isospin violation, the EM decay width may be less different from the strong decay width.

There have been discussions of its strong and EM decays using different models \cite{PRD68,PLB566,PLA19,PLB568,PRD69,PRD70,PLB570}. Among them, Godfrey found a large branching ratio to $D_s^*\gamma$ for $D_{s0}^*(2317)$; Bardeen $et~al$ \cite{PRD68} studied $D_s+\pi$, $D_s+2\pi$, and EM transitions from the full chiral theory; Colangelo and Fazio used the method based on heavy quark symmetries and the vector meson dominance ansatz to study the strong and EM decays; Zhu $et~al$ used the $3P_0$ model \cite{PRD73} and the light-cone QCD sum rule (LCQSR) \cite{PRD73034004} to study the strong decays of $D_{s0}^*(2317)$; Liu $et~al$ studied the strong pionic and radiative decays in the Constituent Quark Meson model \cite{PJC47}; Wang calculated the corresponding strong coupling constant in the framework of the LCQSR \cite{PRD75} and the radiative decay with the assumption of vector meson dominance; Guo $et~al$ studied the strong decay by constructing an effective chiral Lagrangian under the assumption of a hadronic molecule \cite{PLB666}; and Liu $et~al$ \cite{Liu:2012zya} studied the strong decay under the assumption of a $DK$ molecule in lattice QCD.

In these studies, there is a lack of the relativistic method. Therefore, in this paper, we assume $D_{s0}^*(2317)$ as the $c\bar{s}$ traditional meson and study its strong and EM decays under the Bethe-Salpeter (BS) framework. The BS equation is a formally exact equation to describe the relativistic bound state \cite{PR84,PR87}. Using the BS method, we have studied the strong decays \cite{EPJC2018,JHEP} and electromagnetic decays \cite{PLB697,PG40} of some particles, and the results are in good agreement with the experimental data. In this paper, we will show the advantages of the relativistic BS method, which can help reveal why the decay width of $D_{s0}^*(2317)$ is so small, especially the radiative electromagnetic decay width.

The remainder of this paper is organized as follows. We show the hadronic matrix elements and the formula for the two-body strong decay width of  $D_{s0}^*(2317) \to D_s\pi^0$ through $\eta-\pi^0$ mixing in Section \uppercase\expandafter{\romannumeral2}. EM decay is presented in Section \uppercase\expandafter{\romannumeral3}. We show and compare our results with experimental data and other theoretical approaches in literature and summarize our results in Section \uppercase\expandafter{\romannumeral4}.

\section{two-body strong decay of $D_{s0}^*(2317)$ }
In this section, we give the formula for calculating the hadronic transition matrix element and the strong decay, as well as the relativistic wave functions used.
\begin{figure}[h]
\centering
\includegraphics[width=0.6\textwidth]{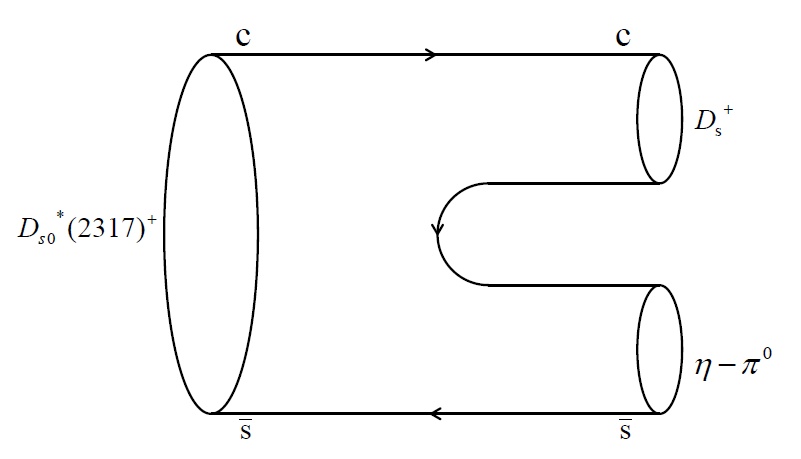}
\vspace{0pt}
\setlength{\belowcaptionskip}{0cm}
\caption{Original strong decay diagram for $D_{s0}^*(2317)^+\to D_{s}^+\pi^0$.}
\label{Fig.1}
\end{figure}

\begin{figure}[h]
\centering
\includegraphics[width=0.7\textwidth]{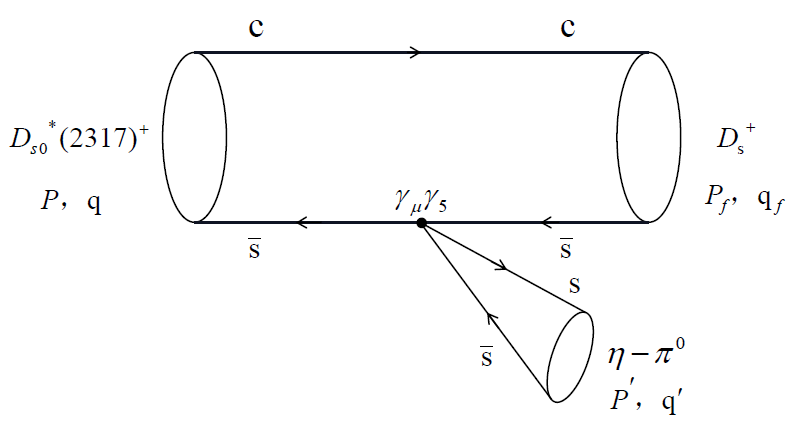}
\caption{Strong decay diagram for $D_{s0}^*(2317)^+\to D_{s}^+\pi^0$ after approximation. }
\label{Fig.2}
\end{figure}

\subsection{\small Hadronic transition matrix element}
When both quarks in the initial state appear in the final state meson, the strong decay of $D_{s0}^*(2317)^+$ can be expressed by a Feynman diagram, as shown in Fig.\ref{Fig.1}.
The transition matrix element of the two-body strong decay $D_{s0}^{*}(2317)^{+ }\to D_{s}^{+}\eta $ can be expressed as (we compute the $D_{s0}^{*}(2317)^{+}\to D_{s}^{+}\eta $ process first, and then we import $\eta-\pi^0$ mixing for $D_{s}^{+}\eta \to D_{s}^{+}\pi ^{0}$)\cite{PLB623}
\begin{equation}
\left< D_{s}^{+}(P_{_f})\eta (P^{'})| D_{s0}^{*+}(P)\right> =\int d^{4}xe^{iP^{'}x}(M_{\eta }^{2}-P^{'2})\left<  D_{s}^{+}(P_{_f})|\Phi _{\eta }(x)|D_{s0}^{*+}(P)\right>\label{amplitude},
\end{equation}

\noindent where, $P$, $P_{_f}$, and $P^{'}$ represent the momenta of  $D_{s0}^*(2317)^+$, $D_{s}^{+}$ and $\eta$, respectively. $\Phi _{\eta }(x)$ represents the field of $\eta$. And the light meson field is expressed as the derivative of the axial-vector flow divided by the decay constant $f_{_P}$ of the light pseudoscalar meson according to the PCAC relation:
\begin{equation}
\Phi_{\eta } (x)=\frac{1}{M_{\eta }^{2}f_{_P }}\partial ^{\mu }(\bar{s}\gamma _{\mu }\gamma _{5}s)\label{relation}.
\end{equation}
Fig.\ref{Fig.1} can be approximately converted into Fig.\ref{Fig.2} by using Eq.\ref{amplitude}.

By combining Eqs.(\ref{amplitude}) and (\ref{relation}), we obtain
\begin{equation}
\left< D_{s}^{+}(P_{_f})\eta (P^{'})| D_{s0}^{*+}(P)\right> =\frac{M_{\eta }^{2}-P^{'2}}{M_{\eta }^{2}f_{_P}}\int d^{4}xe^{iP^{'}x}\left<D_{s} ^{+}(P_{_f})|\partial ^{\mu }\left ( \bar{s}\gamma _{\mu }\gamma _{5}s \right )|D_{s0}^{* +}(P)\right>.
\end{equation}
By using partial integration and applying the low-energy theorem, we can obtain the form of the transition matrix element in the momentum space. It can be written as
\begin{equation}
\begin{aligned}
\left< D_{s}^{+}(P_{_f})\eta (P^{'})| D_{s0}^{*}(P)\right> &=\frac{M_{\eta }^{2}-P^{'2}}{M_{\eta }^{2}f_{_P }}\int d^{4}xe^{iP^{'}x}\left<D_{s} ^{+}(P_{_f})|\partial ^{\mu }\left ( \bar{s}\gamma _{\mu }\gamma _{5}s \right )|D_{s0}^{* +}(P)\right>
\\&=\frac{-iP^{'\mu }\left ( M_{\eta }^{2}- P^{'2}\right )}{M_{\eta }^{2}f_{_P }}\int d^{4}xe^{iP^{'}x}\left<D_{s} ^{+}(P_{_f})| \bar{s}\gamma _{\mu }\gamma _{5}s |D_{s0}^{*+}(P)\right>
\\&\approx (2\pi )^{4}\delta ^{4} (P-P_{_f}-P^{'} )\frac{-iP^{'\mu }}{f_{_P }}\left<D_{s} ^{+}(P_{_f})| \bar{s}\gamma _{\mu }\gamma _{5}s |D_{s0}^{*+}(P)\right>.
\end{aligned}
\end{equation}
Then, the transition amplitude for the process $D_{s0}^{*}(2317)^{+}\to D_{s}^{+}+\eta$ is
\begin{equation}
T=\frac{-iP^{'\mu }}{f_{_P }}\left<D_{s} ^{+}(P_{_f})| \bar{s}\gamma _{\mu }\gamma _{5}s |D_{s0}^{*+}(P)\right>.
\end{equation}

In the Mandelstam form \cite{PRL233}, the transition amplitude can be written as an overlapping integral of the initial and final meson wave functions \cite{TP46}
\begin{equation}
\begin{aligned}
\mathcal{M}&=\left<D_{s} ^{+}(P_{_f})| \bar{s}\gamma _{\mu }\gamma _{5}s |D_{s0}^{*+}(P)\right>\\
&=\int  \frac{d^{4}q}{\left ( 2\pi  \right )^{4}}\frac{d^{4}q_{_f}}{\left ( 2\pi  \right )^{4}}Tr\left [\bar{\chi }_{P_{_f}}\left ( q_{_f} \right )S_{1}^{-1}\chi _{P} \left ( q \right )\gamma _{\mu }\gamma _{5 }\right ]\left ( 2\pi  \right )^{4}\delta ^{4}\left ( p_{_1}-p_{_{1f}} \right )\\
&\approx\int\frac{d^{3}q_{\perp }}{2\pi ^{3}}Tr\left[\bar{\varphi }_{P_{_f}}^{++}(q_{\perp }-\alpha _{1}P_{_{f\perp }}) \frac{\slashed P}{M}\varphi _{P}^{++}\left(q_{\perp }\right)\gamma _{\mu }\gamma _{5}\right],
\end{aligned}
\end{equation}
where $\chi _{_P} \left ( q \right )$ and ${\chi }_{_{P_{_f}}}\left ( q_{_f} \right )$ represent the relativistic BS wave functions of $D_{s0}^{*}(2317)^{+}$ and $D_{s}^{+}$, respectively. $ q$ and $q_{_f}$ represent the internal relative momenta of the initial and final mesons, respectively. $p_{_1}$, $p_{_2}$, $p_{_{1f}}$ and $p_{_{2f}}$ represent the momenta of the quarks and anti-quarks of the initial and final mesons, respectively. $S_{1}$ represents the propagator of the quarks. $M$ represents the mass of $D_{s0}^{*}(2317)^{+ }$, and $\alpha_{1} =\frac{m_{c}}{m_{c}+m_{s}}$ with the quark mass $m_{c}$ and the anti-quark mass $m_{\bar{s}}$; $\varphi _{P_{_f}}^{++}$ and $\varphi _{P}^{++}$ represent the positive-energy wave functions of $D_{s}^{+}$ and $D_{s0}^{*}(2317)^{+ }$, respectively. $P_{_{f\perp} }$ is defined as $P_{_{f\perp} }^{\mu }=P_{_f}^{\mu }-(\frac{P\cdot P_{_f}}{M^{2}})P^{\mu }$ (similar definition to $q_{\perp}$), and  $\bar{\varphi }_{P_{_f }}^{++}$ is defined as $\gamma_{0}( \varphi _{P_{_f }}^{++})^{\dagger }\gamma _{0}$.

\subsection{\small Relativistic wave functions}
In our model, we use the modified BS method based on the constituent quark model to give the wave functions for $J^{P}$ or $J^{PC}$ states. Here, we will directly give the corresponding positive energy wave functions of initial and final states.
\subsubsection{$0^{+}$ state $D_{s 0}^{*+}$}
The relativistic positive energy Salpeter wave function for the $J^P$ = $0^+$ state $D_{s0}^{*}(2317)^{+}$ ($^3P_0$ state), can be written as
\begin{equation}
\varphi _{0^{+}}^{++}\left (P, q_{\perp } \right )=A_{1}\slashed q_{\perp }+A_{2}\frac{\slashed P\slashed q_{\perp }}{M}+A_{3}+A_{4}\frac{\slashed P}{M}:
\end{equation}
where the $A_{i}$ (i=1,2,3,4) are related to the original radial wave functions $f_{1}$ and $f_2$ of the $0^{+}$ wave function \cite{PLB706}, quark mass $m_{i}$, quark energy $\omega_{i}$ (i=c,s), and meson mass $M$,
\[A_{1}=\frac{1}{2} ( f_{1}+f_{2} \frac{m_{c}+m_{s}}{\omega _{c}+\omega _{s}} ),~~~A_{2}=\frac{\omega _{c}+\omega _{s}}{m_{c}+m_{s}}A_{1},\]
\[A_{3}=q_{\perp }^{2}\frac{\omega _{c}+\omega _{s}}{m_{c}\omega _{s}+m_{s}\omega _{c}}A_{1},~~~A_{4}=\frac{m_{s}\omega _{c}-m_{c}\omega _{s}}{m_{c}+m _{s}}A_{1}.\]

We should note that the wave function of a meson is not a pure wave; it actually includes different partial waves. For example, in the positive energy wave function, the  $A_1$ and $A_2$ terms are $P$ $waves$, which provide the main non-relativistic contribution, while the $A_3$ and $A_4$ terms are $S$ $waves$, which give the main relativistic corrections \cite{JHEP05}.
\subsubsection{$0^{-}$ state $D_{s}^{+}$}
The relativistic positive energy Salpeter wave function for $D_{s}^{+}$ ($^1S_0$ state) with $J^P$ = $0^-$ can be written as
\begin{equation}
\varphi_{0^{-}}^{++} \left (P, q_{\perp } \right )=\left [ B_{1}\frac{\slashed P}{M}+B_{2}+B_{3}\slashed q_{\perp }+B_{4}\frac{\slashed q_{\perp }\slashed P}{M}\right ]\gamma _{5},
\end{equation}
where the parameters $B_{i}$ are functions of the radial wave functions $g_{1}$ and $g_{2}$ for the $0^{-}$ state \cite{PLB706}:
\[ B_{1}=\frac{M}{2} ( g_{1}+g_{2}\frac{m_{c}+m_{s}}{\omega _{c}+\omega _{s}}  ),~~~B_{2}=\frac{\omega _{c}+\omega _{s}}{m_{c}+m_{s}}B_{1},\]
\[B_{3}=-\frac{m_{c}-m_{s}}{m_{c}\omega _{s}+m_{s}\omega _{c}}B_{1},~~~B_{4}=\frac{\omega _{c}+\omega _{s}}{m_{c}\omega _{s}+m_{s}\omega _{c}}B_{1}.\]

We point out that the $B_1$ and $B_2$ terms are $S$ $waves$, which are the main partial waves of the meson, and the $B_3$ and $B_4$ terms are $P$ $waves$ which provide the main relativistic corrections \cite{JHEP05}.
\subsection{\small Decay width of strong decay}
The two-body decay width formula can be expressed (import $\eta-\pi^0$ mixing) as
\begin{equation}
\Gamma =\frac{1}{8\pi}\frac{\left|\vec{P_{_f}}\right|}{M^{2}}\frac{1}{2J+1}\sum_{\lambda }\left|\frac{Tt_{\pi \eta }}{m^2_{\pi }-m^2_{\eta }}\right|^{2},
\end{equation}
where $|\vec{P_{_f}}|=\sqrt{[M^{2}-(M_{f}-m_{\pi })^{2}][M^{2}-(M_{f}+m_{\pi })^{2}]}/(2M)$, which is the three-momenta of the final meson $D_{s}^{+}$. $M$, $M_{f}$, and $m_{\pi }$ represent the masses of the initial state and two final states, respectively. $J$ represents the spin quantum number of the initial meson, which is 0 in this case. $t_{\pi \eta }$ is the mixing matrix entry of $\eta-\pi^0$, and $t_{\pi \eta }=\left< \pi^{0}\left| \mathcal{H}\right|\eta \right>=-0.003$ GeV$^2$ \cite{PR1969} is chosen.
\section{EM decay of $D_{s0}^*(2317)^+$ }
Similarly, we give the Feynman diagram of the EM decay of $D_{s0}^*(2317)^+$, corresponding to the two subplots of Fig.\ref{Fig.3}:

\begin{figure}[htb]
\centering
\includegraphics[width=0.9\textwidth]{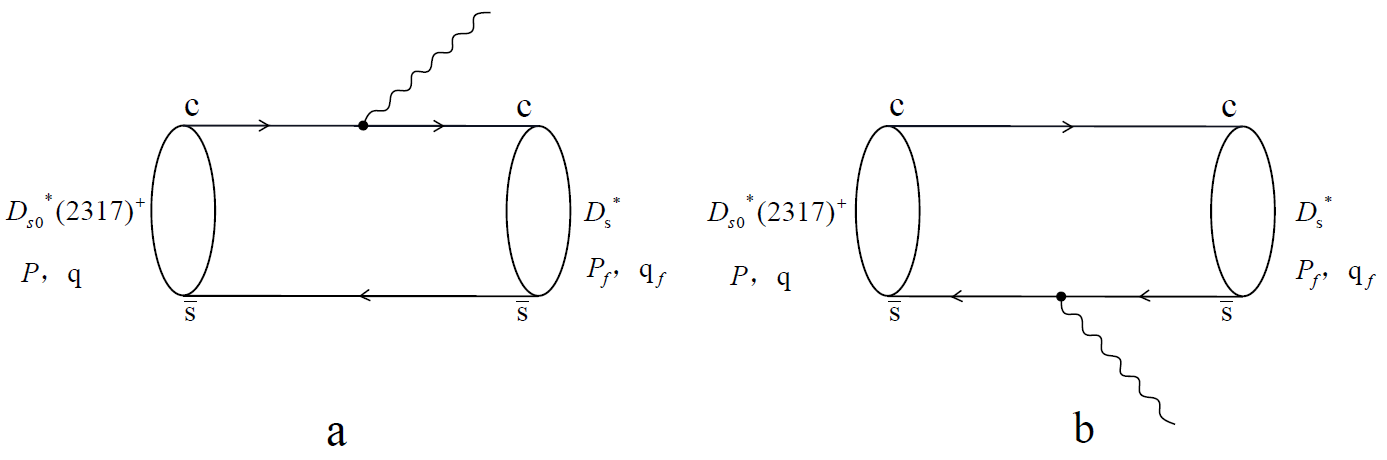}
\caption{Feynman diagram responsible for the EM decay $D_{s0}^*(2317)^+\to D_{s}^{*+}\gamma$ }
\label{Fig.3}
\end{figure}

\subsection{EM transition matrix element}
The transition matrix element of the EM decay $D_{s0}^*(2317)^+\rightarrow D_{s}^{*}\gamma $ can be written as
\begin{equation}
\left< D_{s}^{*}\left ( P_{_f},\epsilon\right ) \gamma \left ( k,\epsilon _{0} \right )| D_{s0}^{*+}\right>=(2\pi )^{4}\delta ^{4}\left (  P-P_{_f}-k\right )\epsilon _{0\mu }\mathcal{M}^{\mu },
\end{equation}
where $\epsilon _{0}$ and $\epsilon$ represent the polarization vectors of the photon and final meson, respectively. 
 $P$, $P_{_f}$, and $k$ represent the momenta of the initial meson, final meson, and photon, respectively.

$\mathcal{M}$ represents the invariant amplitude, which is expresed as
\begin{equation}
\begin{aligned}
\mathcal{M}^{\mu }=\int \frac{d^{4}q}{\left ( 2\pi  \right )^{4}}\frac{d^{4}q_{_f}}{\left ( 2\pi  \right )^{4}}Tr&\left[\bar{\chi }_{P_{_f}}\left ( q_{_f} \right )Q_{1}\gamma ^{\mu }\chi _{P} \left ( q \right )\left ( 2\pi  \right )^{4}\delta ^{4}( p_{2}-p_{2}^{'}  )S_{2}^{-1}\left ( p_{2} \right )\right.\\
&\left.+\bar{\chi }_{P_{_f}}\left ( q_{_f} \right )\left ( 2\pi  \right )^{4}\delta ^{4}( p_{1}-p_{1}^{'} )S_{1}^{-1}\left ( p_{1} \right )\chi _{P} \left ( q \right )Q_{2}\gamma ^{\mu }\right ].
\end{aligned}
\end{equation}
Here, $Q_{1}$ and $Q_{2}$ represent the electric charges (in unit of $e$) of quark and anti-quark, respectively. We use the instantaneous approximation to obtain the following form of the amplitude \cite{TP46}:
\begin{equation}
\begin{aligned}
\mathcal{M}^{\mu }=\int \frac{d^{3}q}{\left ( 2\pi  \right )^{3}}Tr&\left[Q_{1}\bar{\varphi  }_{P_{_f}}^{++}\left ( q_{\perp }+\alpha _{2}P_{f\perp } \right )\gamma ^{\mu }\varphi _{P}^{++} \left ( q_{\perp } \right )\frac{\slashed P}{M}\right.\\
&\left.+Q_{2}\bar{\varphi  }_{P_{_f}}^{++}\left ( q_{\perp } -\alpha _{1}P_{f\perp }\right )\frac{\slashed P}{M}\varphi  _{P} ^{++}\gamma ^{\mu }\right ],
\end{aligned}
\end{equation}
where $M$ represents the mass of $D_{s0}^{*}(2317)^{+ }$, and $q_{f\perp }=q_{\perp }+\alpha _{2}P_{f\perp }$ and $q_{f\perp }=q_{\perp }-\alpha _{1}P_{f\perp }$ correspond to two different processes. $\alpha_{1} =\frac{m_{c}}{m_{c}+m_{s}}$ and $\alpha_{2} =\frac{m_{s}}{m_{c}+m_{s}}$ are substituted into the above formula with the quark mass $m_{c}$ and the anti-quark mass $m_{\bar{s}}$, respectively.
\subsection{Relativistic wave function for $1^{-}$ state $D_{s}^{*}$}
The relativistic positive energy Salpeter wave function for $D_{s}^{*}$ ($^3S_1$ state) can be written as
\begin{equation}
\begin{aligned}
\varphi _{1^{-}}^{++}\left (P,q_{\perp } \right )=&C_{1}\slashed \varepsilon +C_{2}\slashed \varepsilon \slashed P+C_{3}\left ( \slashed q_{\perp } \slashed \varepsilon -q_{\perp }\cdot \varepsilon \right )+C_{4}\left ( \slashed P\slashed \varepsilon \slashed q_{\perp } -\slashed Pq_{\perp }\cdot \varepsilon \right )\\
&+q_{\perp }\cdot \varepsilon \left (C_{5} +C_{6}\slashed P+C_{7} \slashed q_{\perp }+C_{8}\slashed q_{\perp }\slashed P\right ),
\end{aligned}
\end{equation}
where we define the parameters $C_{i}$ using the radial wave functions $f_{i}$ (i=3,4,5,6) for a $1^-$ state \cite{PLB706}:
\begin{align*}
&C_{1}=\frac{M}{2}\left ( f_{5} -f_{6}\frac{\omega _{c}+\omega_{s} }{m_{c}+m_{s}}\right ),\\
&C_{2}=-\frac{1}{2} \frac{m_{c}+m_{s}}{\omega _{c}+\omega _{s}} \left ( f_{5} -f_{6}\frac{\omega _{c}+\omega_{s} }{m_{c}+m_{s}}\right ),\\
&C_{3}=\frac{M}{2} \frac{\omega _{s}-\omega _{c}}{m_{c}\omega _{s}+m_{s}\omega _{c}} \left ( f_{5} -f_{6}\frac{\omega _{c}+\omega_{s} }{m_{c}+m_{s}}\right ),\\
&C_{4}=\frac{1}{2} \frac{\omega _{c}+\omega _{s}}{\omega _{c}\omega _{s}+m_{c}m_{s}-q_{\perp }^{2}} \left ( f_{5} -f_{6}\frac{\omega _{c}+\omega_{s} }{m_{c}+m_{s}}\right ),\\
&C_{5}=\frac{1}{2M} \frac{m _{c}+m _{s}}{\omega _{c}\omega _{s}+m_{c}m_{s}+q_{\perp }^{2}}\left[M^{2}\left ( f_{5} -f_{6}\frac{m_{c}+m_{s} }{\omega _{c}+\omega_{s}}\right )+q_{\perp }^{2}\left ( f_{3} +f_{4}\frac{m_{c}+m_{s}}{\omega _{c}+\omega_{s}}\right )\right],\\
&C_{6}=\frac{1}{2M^{2}} \frac{\omega  _{c}+\omega  _{s}}{\omega _{c}\omega _{s}+m_{c}m_{s}+q_{\perp }^{2}}\left[M^{2}\left ( f_{5} -f_{6}\frac{m_{c}+m_{s} }{\omega _{c}+\omega_{s}}\right )+q_{\perp }^{2}\left ( f_{3} +f_{4}\frac{m_{c}+m_{s}}{\omega _{c}+\omega_{s}}\right )\right],\\
&C_{7}=\frac{1}{2M}\left(f_{3} +f_{4}\frac{m_{c}+m_{s}}{\omega _{c}+\omega_{s}}\right)-\frac{f_{6}M}{m_{c}\omega _{s}+m_{s}\omega _{c}},\\
&C_{8}=\frac{1}{2M^{2}}\frac{\omega _{c}+\omega_{s}}{m_{c}+m_{s}} \left(f_{3} +f_{4}\frac{m_{c}+m_{s}}{\omega _{c}+\omega_{s}}\right)-f_{5}\frac{\omega _{c}+\omega_{s}}{\left ( m_{c}+m_{s} \right )\left ( \omega _{c}\omega _{s}+m_{c}m_{s}-q_{\perp }^{2} \right )}.
\end{align*}

The positive energy wave function $\varphi _{1^{-}}^{++}\left (P,q_{\perp } \right )$ for $D_{s}^{*}$ includes three partial waves: the  $C_1$ and $C_2$ terms are $S$ $waves$; the $C_3$, $C_4$, $C_5$, and $C_6$ terms are $P$ $waves$; and the $C_7$ and $C_8$ terms are $D$ $waves$ \cite{JHEP05}.


\section{Result and discussion}
In this paper, the following masses of constituent quarks are adopted: $m_{c}$ = 1.62 GeV and $m_{s}$ = 0.50 GeV. We take the meson mass to have the following values: $M_{D_{s0}^{*+}}$ = 2.317 GeV, $M_{D_{s}^{+}}$ = 1.968 GeV, $M_{D_{s}^*}$ = 2.112 GeV, $M_{\eta}$ = 0.548 GeV, and $M_{\pi^0}$ = 0.135 GeV\cite{PTEP}. The decay constant is $f_{\pi}$ = 0.130 GeV.
\subsection{ Numerical results for decay widths}
\subsubsection{Strong decay}
Our relativistic prediction result for the strong decay of $D_{s0}^{*}(2317)^{+ }$ is
\begin{equation}
\Gamma(D_{s0}^{*+}\to D_s^+\pi^0) = 7.83_{-1.55}^{+1.97}~ {\rm keV},
\end{equation}
where the theoretical uncertainties are also shown. These uncertainties are calculated by varying all the input parameters simultaneously within $\pm 5\%$ of the central value, and the largest variation is taken.

For comparison, we list our prediction and those of other groups in Table \ref{tab:strong decay compare}. As shown, our prediction $ \Gamma(D_s^+\pi^0)$ = $7.83_{-1.55}^{+1.97}$ keV is consistent with the theoretical results in Refs. \cite{PLB570,PJC47,PRD68,PLB634}. Among them, Ref.\cite{PLB570} is based on the heavy quark symmetries and vector meson dominance ansatz; Ref.\cite{PJC47} is based on the framework of the Constituent-Quark-Meson (CQM) model; and Ref.\cite{PRD68} is based on full chiral theory. We also refer to Ref.\cite{PLB634}, in which the QCD sum rules were used to analyze the strong decay process under the hypothesis of the four-quark state.  In Table \ref{tab:strong decay compare}, the last two predictions of Refs.\cite{PLB666,Liu:2012zya}, which assume that $D_{s0}^{*}(2317)^{+ }$ is a $DK$ hadronic molecule, differ significantly from our prediction and others,indicating that the strong decay may be crucial to test the nature of $D_{s0}^{*}(2317)^{+ }$.

\begin{table}[htb]
\centering
\caption{Strong decay widths (in units of keV) of $D_{s0}^{*}(2317)^{+ }$}\label{tab:strong decay compare}
\begin{tabular}{c|c|c|c|c|c|c|c|c}
\hline
      Decay chnnal   &Our work&\cite{PLB570}&\cite{PJC47}&\cite{PLB568}&\cite{PRD68}&\cite{PLB634}
      &\cite{PLB666}&\cite{Liu:2012zya}\\
\hline
$\Gamma(D_{s0}^{*+}\to D_{s}^+\pi^{0})$& $7.83_{-1.55}^{+1.97}$  &     6      & 3.68-8.71  & $ \sim $10  &     7.74    & 6$\pm$2 &$180\pm110$&$133\pm22$ \\
\hline
\end{tabular}
\end{table}

\subsubsection{EM decay}
Our relativistic prediction of the EM decay of $D_{s0}^{*}(2317)^{+ }$ is
\begin{equation}
\Gamma(D_{s0}^{*+}\to D_s^*\gamma) = 2.55_{-0.45}^{+0.37} ~{\rm keV}.
\end{equation}
Our result and those from other theoretical groups are presented in Table \ref{tab:EM decay compare} for comparison. Our result is slightly larger than those from Refs.\cite{PLB570,PJC47,PLB568} (within a reasonable margin of error); consistent with the prediction in Ref.\cite{PRD75}, which was calculated in the framework of the light-cone QCD sum rules, and slightly smaller than the prediction in Ref.\cite{fu2022}, which assumes that $D_{s0}^{*}(2317)^{+ }$ is a $DK$ molecule.
\begin{table}[htb]
\centering
\caption{EM decay widths (in unit of keV) of $D_{s0}^{*}(2317)^{+ }$}\label{tab:EM decay compare}
\begin{tabular}{c|c|c|c|c|c|c}
\hline
       Decay chnnal                  &our work&\cite{PLB570}&\cite{PJC47}&\cite{PLB568}&\cite{PRD75}&\cite{fu2022}\\
\hline
$\Gamma(D_{s0}^{*+}\to D_{s}^*\gamma)$&  $2.55_{-0.45}^{+0.37}$   & 1  & $\sim$1.1  &    1.9    &  1.3-9.9 & $3.7\pm 0.3$ \\
\hline

\end{tabular}
\end{table}

\subsection{Contributions of different partial waves}
As mentioned previously, the complete relativistic wave function for a $J^P$ state is not a pure wave. It includes different partial waves \cite{JHEP05}: both the non-relativistic main part and the relativistic correction terms are included.
We calculate the contributions (to the decay width) of different partial waves in the initial and final states. The results for strong decay are presented in Table \ref{tab:partical waves for SD}, and those for EM decay are presented in Table \ref{tab:partical waves for EMD}. In the tables, ``$complete$" means that the complete wave function is used, ``$S$ $wave$" means that only the $S$ partial wave makes a contribution and other partial waves are ignored.
\subsubsection{Strong decay}

\begin{table}[htb]
\centering
\caption{Strong decay width (keV) of different partial waves for $D_{s0}^{*+}\to D_{s}^+\pi^{0}$ $(0^+\to0^-)$}\label{tab:partical waves for SD}
\begin{tabular}{c|c|c|c}
\hline
\diagbox{$0^+$}{$0^-$}&       $complete$        &   $S$ $wave$ $(B_1,B_2)$   &  $P$ $wave$ $(B_3,B_4)$        \\
\hline
$complete$            & $7.83_{-1.55}^{+1.97}$  &    $16.9_{-2.11}^{+2.52}$  &  $1.74_{-0.09}^{+0.08}$        \\
\hline
$P$ $wave(A_1,A_2)$   & $11.3_{-1.95}^{+2.43}$  &    $21.8_{-2.50}^{+2.94}$  &  $1.70_{-0.10}^{+0.08}$        \\
\hline
$S$ $wave(A_3,A_4))$  &     $0.32\pm0.01$       &       $ 0.30\pm0.01$       &  $2.27_{-0.03}^{+0.04}\times 10^{-4}$         \\
\hline
\end{tabular}
\end{table}

From Table \ref{tab:partical waves for SD}, we can see that the main contribution of the $0^+$ state $D_{s0}^{*+}$ comes from its $P$ $wave$, which is the non-relativistic part, and the relativistic correction ($S$ $wave$) makes a small contribution. Meanwhile, for the $0^-$ state $D_{s}^+$, the main contribution comes from its $S$ partial wave, which is the non-relativistic part, and the relativistic correction part ($P$ partial wave) makes a very small contribution.

In the strong decay, If we only keep the non-relativistic wave functions and ignore the other partial waves for both the initial and final states, we obtain the non-relativistic result
 \begin{equation}
  \Gamma_0(D_{s0}^{*+}\to D_s^+\pi^0) = 21.8_{-2.50}^{+2.94}~ {\rm keV},
  \end{equation}
which is far larger than the relativistic one of $\Gamma = 7.83^{+1.97}_{-1.55}$ keV, indicating that the relativistic correction in this transition is very large.
\subsubsection{EM decay}

\begin{table}[htb]
\centering
\caption{ EM decay width (keV) of different partial waves for $D_{s0}^{*+}\to D_{s}^*\gamma$ $(0^+\to1^-)$}\label{tab:partical waves for EMD}
\begin{tabular}{c|c|c|c|c}
\hline
\diagbox{$0^+$}{$1^-$}& $complete$ & $S$ $wave$ $(C_1,C_2)$&$P$ $wave$ $(C_3,C_4,C_5,C_6)$ & $D$ $wave$ $(C_7,C_8)$\\
\hline
$complete$          & $2.55_{-0.45}^{+0.37}$ & $0.16\pm0.05$ & $1.39_{-0.15}^{+0.13}$ & $1.99_{-0.26}^{+0.25}\times 10^{-4}$\\
\hline
$P$ $wave(A_1,A_2)$ & $4.26_{-0.61}^{+0.52}$ & $0.83_{-0.17}^{+0.15}$ & $1.35_{-1.13}^{+0.10}$ &  $7.2_{-2.29}^{+2.46}\times 10^{-5}$        \\
\hline
$S$ $wave(A_3,A_4))$& $0.21\pm0.01$ & $0.26\pm0.02$ & $3.49_{-2.95}^{+4.84}\times 10^{-4}$  &$5.11_{-0.36}^{+1.66}\times 10^{-4}$     \\
\hline
\end{tabular}
\end{table}

Table \ref{tab:partical waves for EMD} presents the contributions of different partial waves in the EM decay $D_{s0}^{*+}\to D_{s}^*\gamma$. The main contribution of $D_{s0}^{*+}$ comes from its $P$ $wave$, which is its dominant partial wave. The final state $D_{s}^*$ is a $1^-$ vector. It is usually denoted as a $^{3}\textrm{S}_{1}$ state in a non-relativistic method, which means that it is an $S$ $wave$ dominated state. We have shown that in addition to the dominant $S$ $wave$, its wave function also includes $P$ and $D$ partial waves, and they provide the relativistic corrections. From Table \ref{tab:partical waves for EMD}, we can see that the contribution of its dominant $S$ partial $wave$ is suppressed, while the $P$ $wave$ gives the main contribution, indicating that the relativistic correction is dominant in this EM decay.
We can see this if we take the non-relativistic limit without relativistic correction. In Table \ref{tab:partical waves for EMD}, if only the $P$ $wave$ in the initial state and the $S$ $wave$ in the final state make contributions, the non-relativistic result is obtained:
\begin{equation}
\Gamma_0(D_{s0}^{*+}\to D_s^*\gamma) = 0.83_{-0.17}^{+0.15}~{\rm keV},
\end{equation}
which is far smaller than the relativistic one of $\Gamma=2.55^{+0.37}_{-0.45}$ keV.

\subsection{Summary}

In this paper, taking the particle $D_{s0}^{*}(2317)^{+ }$ as the conventional $c\bar s$ meson, we calculate its main decay processes in the framework of the Bethe-Salpeter method, $i.e.$, the strong decay and electromagnetic decay. Our results are $\Gamma(D_{s0}^{*+}\to D_s^+\pi^0) = 7.83^{+1.97}_{-1.55}$ keV and $\Gamma(D_{s0}^{*+}\to D_s^{*+}\gamma) = 2.55^{+0.37}_{-0.45}$ keV. Because of the low mass, $D_{s0}^{*}(2317)^{+ }$ has no direct OZI-allowed strong decay channel. Its strong decay channel $D_{s0}^{*}(2317)^{+ }\to D_s^+\pi^0$ violates the isospin conservation, and the $\eta-\pi^0$ mixing is needed. Owing to this mixing effect, the strong decay width is relatively small.

In addition, we calculated the contributions of different partial waves. For the decay of $D_{s0}^{*+}\to D_s^+\pi^0$, the main contribution comes from the dominant partial waves of initial and final states, while for the $D_{s0}^{*+}\to D_s^*\gamma$ decay, the main contribution comes from the dominant partial wave in the initial state and the small $P$ partial wave in the final state. In both cases, the  relativistic corrections are very large.

\section*{Acknowledgements}
This work was supported in part by the National Natural Science Foundation of China (NSFC) under the Grants Nos. 12075073 and 11865001, the Natural Science Foundation of Hebei province under the Grant No. A2021201009, Post-graduate's Innovation Fund Project of Hebei University under the Grant No. HBU2022BS002.

\end{document}